\newcommand{\sonec}{0.47} 
\newcommand{\stwoc}{0.98} 

\def\figa{
  \begin{figure*}[!t] 
    \begin{center} 
      \includegraphics[width=0.85\textwidth]{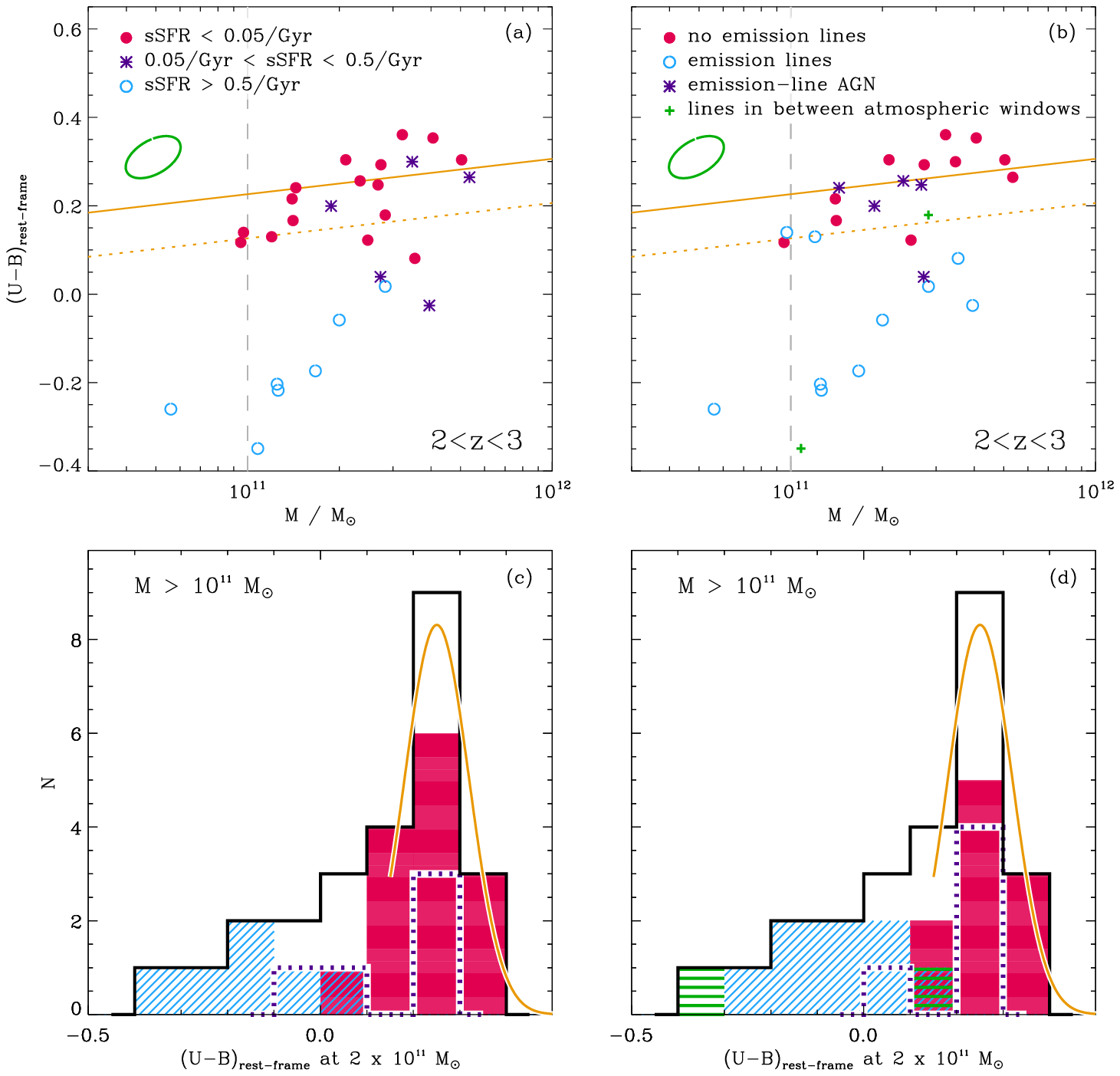} 
      \caption{Rest-frame $U-B$ vs. stellar mass ({\it top panels})
        and the color distribution at $2\times 10^{11} M_{\odot}$
        along the $z\sim0.0$ slope ({\it bottom panels}) for the
        $2<z<3$ massive galaxy sample. In the bottom panels we show
        only galaxies more massive than $10^{11} M_{\odot}$ (to the
        right of the {\em dashed line} in top panels). The black
        histograms in the bottom panels show a significant peak
        ($>3\sigma$), indicating that a red sequence was already in
        place at $z\sim2.3$. The solid curve in the bottom panels
        represents the best fit to the color distribution of the
        red-sequence galaxies. The resulting location of the red
        sequence is indicated by the solid line in the top panels. All
        galaxies above the dotted line in the top panels are defined
        as red-sequence galaxies in this work ($(U-B)_{M} > (U-B)_{\rm
          peak}-0.1$). The left and right panels illustrate the
        properties of the galaxies according to SED modeling and
        emission line diagnostics, respectively. The symbols in panel
        $a$ indicate the best-fit specific SFRs. In panel $b$ the
        different symbols show if emission lines are detected for the
        galaxies, and whether the emission lines are dominated by star
        formation or by AGN activity \citep[see][]{kr07}. The
        corresponding color distributions are presented in panel $c$
        and $d$ by the matching colors. The average 1$\sigma$
        confidence interval is given in the top left of the top
        panels. Both independent star formation indicators imply that
        the red sequence at $z\sim2.3$ is dominated by galaxies with
        little or no ongoing star formation.\label{fig:ub_mass}}
    \end{center} 
  \end{figure*} 
}

\def\figb{
  \begin{figure} \centering
    \includegraphics[width=\sonec\textwidth]{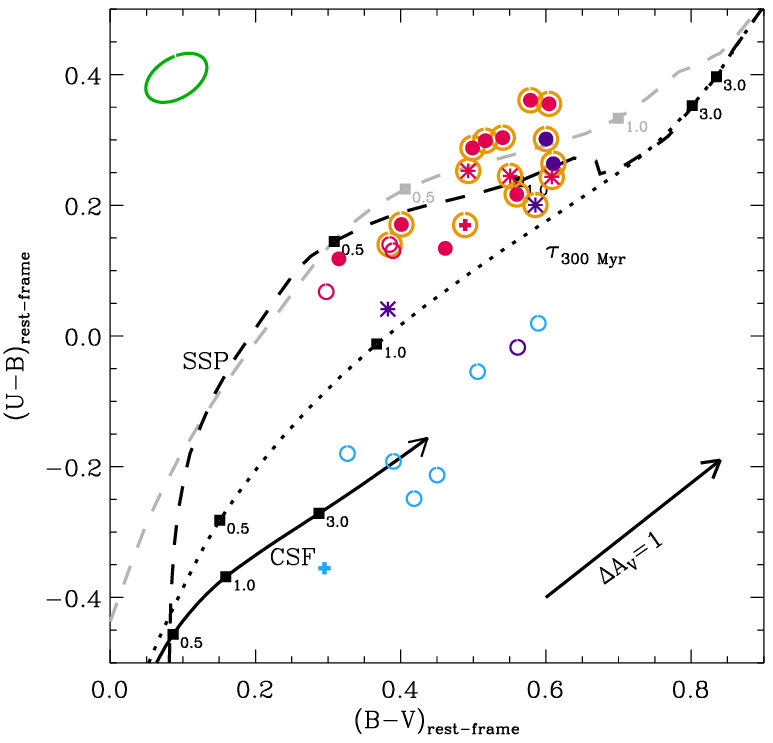}
    \caption{Rest-frame $U-B$ versus $B-V$ for the $2<z<3$ massive
    galaxy sample. The symbols indicate the emission-line diagnostics,
    identical to those in Figure~\ref{fig:ub_mass}b. The colors
    indicate the best-fit specific SFRs similar as the color coding in
    Figure~\ref{fig:ub_mass}a. The red-sequence galaxies are indicated
    by the orange open circles. The average 1\,$\sigma$ confidence
    interval is given in the top left. The black curves show the color
    evolution tracks of \cite{bc03} models for an SSP ({\it dashed
    line}), an exponentially declining model with a $\tau$ of 300 Myr
    ({\it dotted line}), and a CSF model ({\it solid line}), all for
    solar metallicity. The dashed, gray curve represents an SSP model
    with $Z=2.5Z_{\odot}$. Ages in Gyr are indicated along the
    tracks. The vector indicates a reddening of $A_V=1$ mag for a
    \cite{ca00} law. The $B-V$ colors imply that the red-sequence
    galaxies are in a post-starburst phase.\label{fig:ub_bv}}    
  \end{figure} 
}

\def\figc{
  \begin{figure}
    \begin{center} 
      \includegraphics[width=\sonec\textwidth]{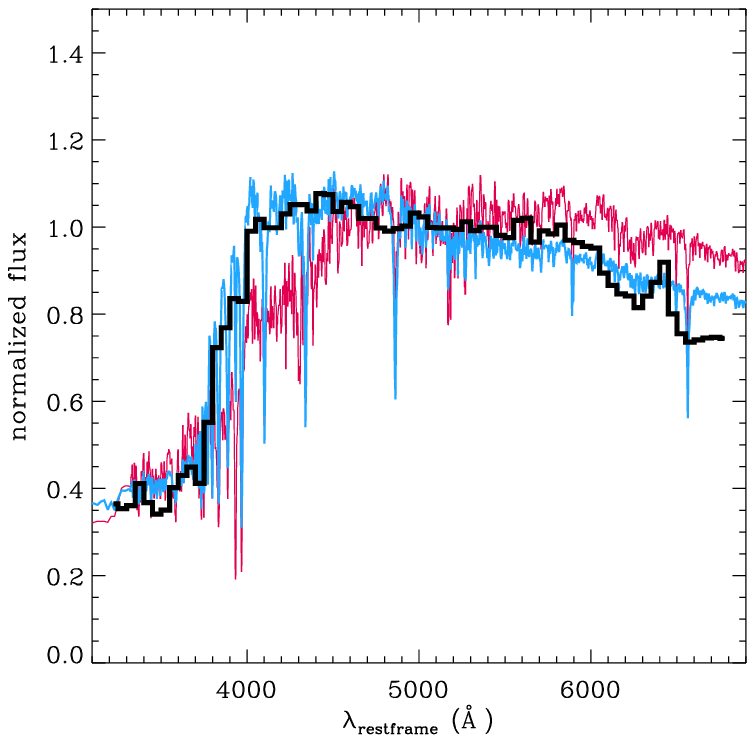} 
      \caption{Mean stack of the low-resolution GNIRS spectra of the
        red-sequence galaxies at $z\sim2.3$ ({\em black
          curve}). Overplotted is the mean of all best fits to the
        spectra in blue. For comparison we show a 2 Gyr SSP model in
        red. This figure shows that in contrast to the 2 Gyr model,
        the optical break of the stacked spectrum is dominated by the
        Balmer break, typical for post-starburst galaxies. This may
        imply that the red sequence has just been starting to build up at
        $z\sim2.3$. \label{fig:stack}}
    \end{center} 
  \end{figure} 
}

  \def\figd{ \begin{figure*}[!t] \begin{center}
    \includegraphics[width=\stwoc\textwidth]{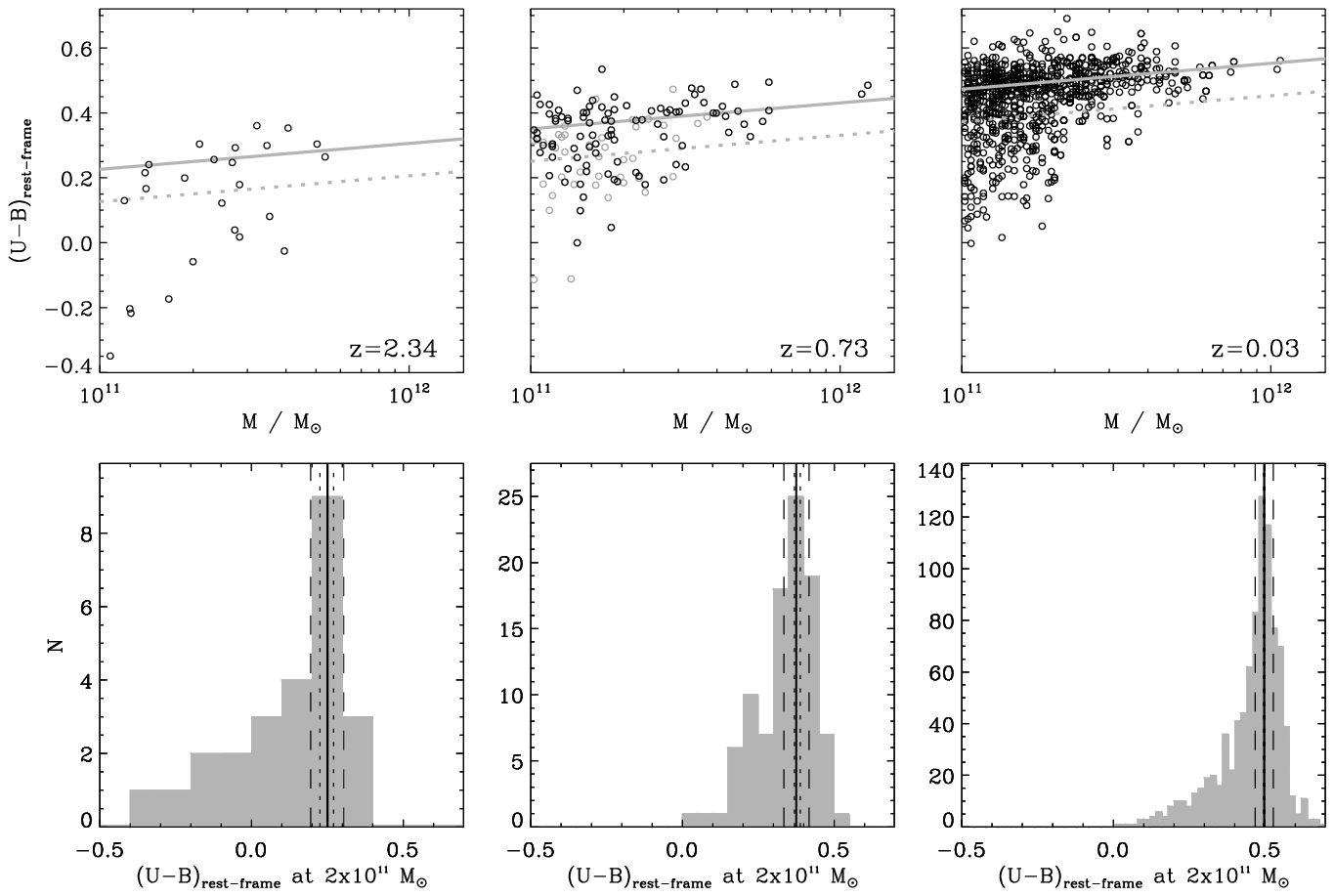} \caption{{\em Top
    panels}: Rest-frame $U-B$ color versus stellar mass for the three
    galaxy samples. The two lower redshift samples are adopted from
    \cite{we07}. Galaxies without spectroscopic redshifts in the
    $0.6<z<1.0$ sample are indicated in gray. The $U-B$ colors are
    corrected for redshift differences within the sample using
    equation~(\ref{eq:ub_t}). {\em Bottom panels}: The color
    distribution extracted along the $z\sim0.03$ slope (0.08 mag
    dex$^{-1}$). The peak of the red sequence is represented by the
    solid gray and black lines in the top and bottom panels
    respectively.  All galaxies above the gray dotted lines in the top
    panels are defined as red-sequence galaxies. The dotted and dashed
    lines in the bottom panels indicate random and total 1$\sigma$
    uncertainties, respectively \label{fig:ub_hist}} \end{center}
    \end{figure*} }

  \def\fige{
  \begin{figure}
    \begin{center} 
      \includegraphics[width=\sonec\textwidth]{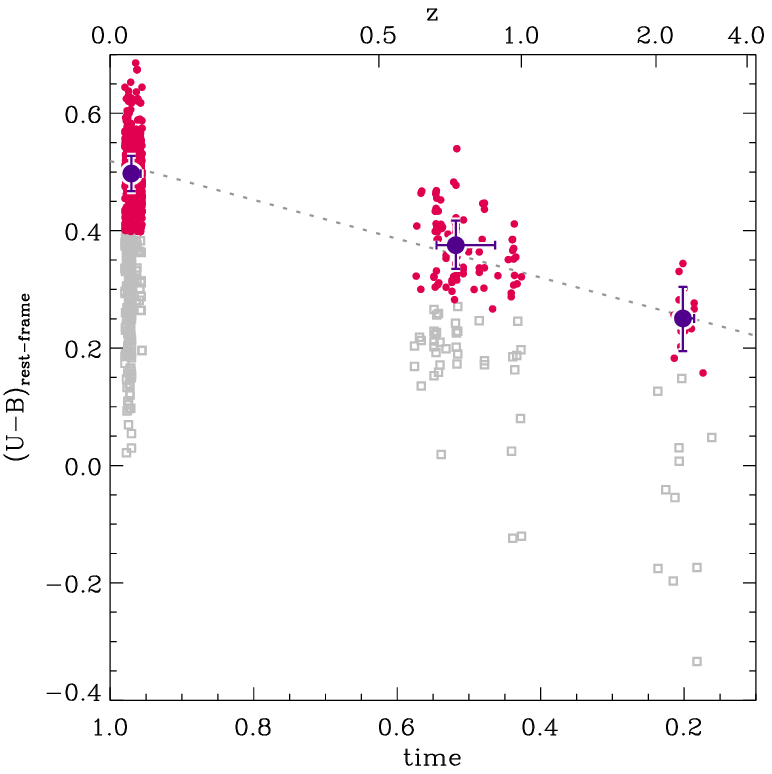} 
      \caption{$U-B$ color versus time for all galaxies in the three
        massive galaxy samples ($>10^{11} M_{\odot}$). The rest-frame
        $U-B$ colors for the individual galaxies in this plot are
        corrected for the slope in the $U-B$ vs. stellar mass relation
        (0.08 mag dex$^{-1}$), and given for a stellar mass of
        $2 \times 10^{11} M_{\odot}$. The red-sequence galaxies are indicated
        by the red filled dots, and the gray open squares represent
        the remaining galaxies. The colors of the red sequence for the
        three samples are indicated by the purple filled symbols. The
        gray dotted line represents the best linear fit through the
        red-sequence locations. This relation is used to correct $U-B$
        colors for redshift differences within the
        samples. \label{fig:ub_t}}
    \end{center} 
  \end{figure} 
}

\def\figf{  
  \begin{figure*}[!t] \centering
    \includegraphics[width=\sonec\textwidth]{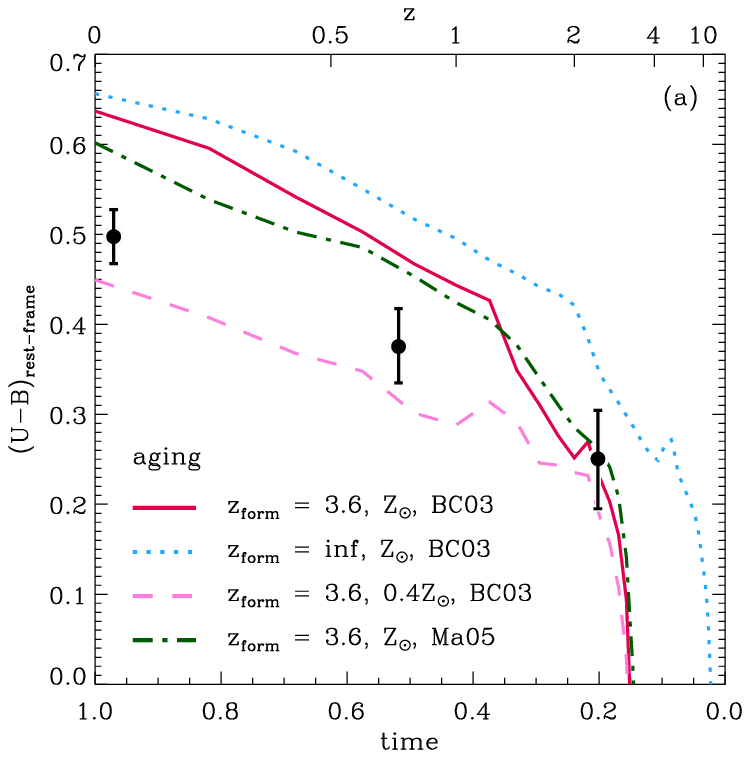} \hspace{-0.4in}
    \includegraphics[width=\sonec\textwidth]{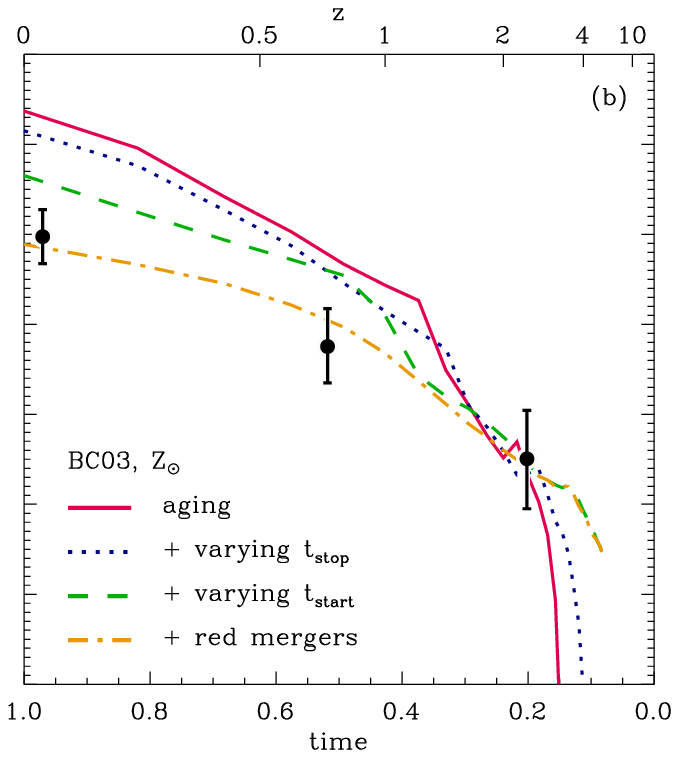}
    \caption{Comparison of the observed rest-frame $U-B$ color
    evolution ({\em black filled circles}) of the red sequence (at $2
    \times 10^{11} M_{\odot}$) with simple models. In panel (a) we
    examine whether the observations are consistent with passive
    evolution. The solid and dashed-dotted lines show the rest-frame
    $U-B$ color evolution for SSP models by \cite{bc03} and
    \cite{ma05}, respectively. Both tracks assume solar metallicity
    and $z_{\rm form}=3.6$. For comparison we also show an SSP model
    for an infinite formation redshift ({\rm dotted curve}), and
    subsolar metallicity ({\rm dashed curve}), both by
    \cite{bc03}. The evolution is reasonably well-fitted by an SSP
    model with subsolar metallicity. However, this is irreconcilable
    with current studies of the metallicity of massive elliptical
    galaxies. In panel (b) we examine more complicated models. The
    dotted curve represents a model in which the red sequence grows by
    newly quenched galaxies. The galaxies have the same $t_{\rm
    start}$, but a different burst length. In the model represented by
    the dashed line we vary $t_{\rm start}$ as well. In the model
    represented by the dashed-dotted line the evolution is further
    flattened due to red mergers. The combination of all effects
    provides a good fit to the observed data. \label{fig:dub_t}}
    \end{figure*} }

\def\figg{  
  \begin{figure}
    \begin{center} 
      \includegraphics[width=\sonec\textwidth]{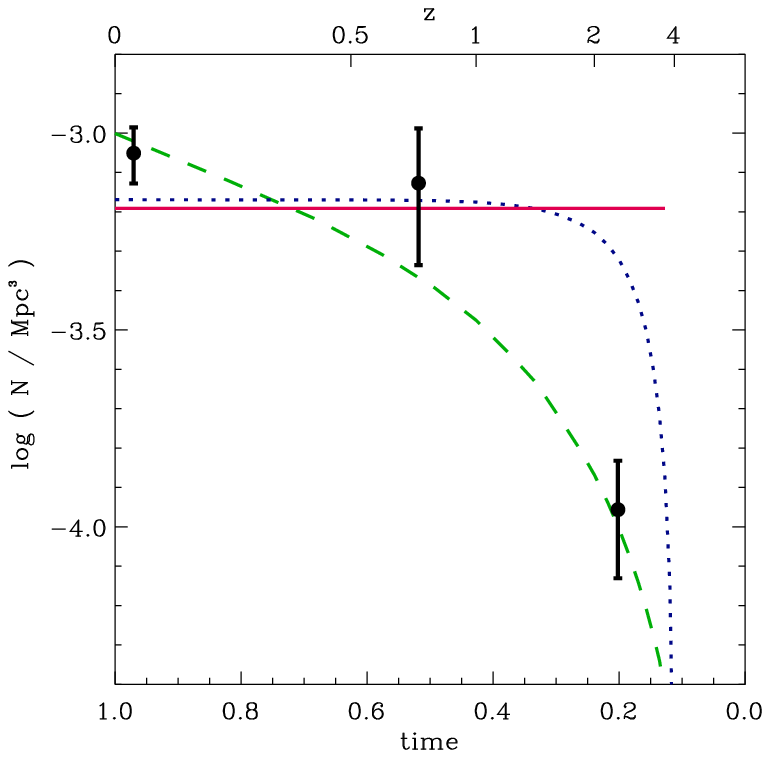} 
      \caption{The evolution of the number density of red-sequence
        galaxies more massive than $10^{11}\,M_{\odot}$ (for a
        Salpeter IMF). The evolution tracks correspond to the models
        in Figure~\ref{fig:dub_t}b. Simple aging models ({\em solid
          curve}) do not match the evolution. The observed evolution
        is well fitted by models with exponentially declining $t_{\rm
          start}$ and $t_{\rm stop}$ (dashed curve: $\tau_{\rm
          start}~\,=\,1$\,Gyr and $\tau_{\rm stop}~\,=\,1$\,Gyr). Also
        models with a large value for $\tau_{\rm stop}$ and a small
        value for $\tau_{\rm form}$, or the other way around, would
        provide good fits. The effect of red mergers on the number
        density evolution are ignored in this
        Figure. \label{fig:dnd_t}}
    \end{center}
  \end{figure}
}

\def\taba{
  \renewcommand{\baselinestretch}{1.4}
  \begin{deluxetable}{l l l l}
    \tabletypesize{\scriptsize}
    \tablecaption{Properties of the Red
      Sequence ($>10^{11} M_{\odot}$)
      \label{tab:rs}}
    \tablewidth{0pt} \tablehead{ & \colhead{$z\sim0.03$} & 
      \colhead{$z\sim0.73$} & \colhead{$z\sim2.34$}}
    \startdata
    Slope (dex$^{-1}$) & 0.08\tablenotemark{a} & 0.08\tablenotemark{b} & 
    0.08\tablenotemark{b}\\
              $U-B$ at $2 \times 10^{11}M_{\odot}$   & $    0.50   ^{+    0.03  
}_{-    0.03}$ & $    0.38   ^{+    0.04  }_{-    0.04}$ & $    0.25   ^{+    0.
06  }_{-    0.06 }$ \\
                                  $\sigma_{(U-B)}$   & $ \hspace{-0.06in}   0.04
8   ^{+   0.002  }_{-   0.002}$ & $ \hspace{-0.06in}   0.055   ^{+   0.011  }_{-
   0.000}$ & $ \hspace{-0.06in}   0.069   ^{+   0.005  }_{-   0.010 }$ \\
                      $N_{\rm RS}~/~N_{\rm total}$   & $    0.77   ^{+    0.02  
}_{-    0.02}$ & $    0.71   ^{+    0.04  }_{-    0.04}$ & $    0.56   ^{+    0.
08  }_{-    0.12 }$ \\
                      $M_{\rm RS}~/~M_{\rm total}$   & $    0.78   ^{+    0.02  
}_{-    0.02}$ & $    0.77   ^{+    0.03  }_{-    0.04}$ & $    0.62   ^{+    0.
06  }_{-    0.11 }$ \\
                   $\rho_N~(10^{-4}~\rm Mpc^{-3})$   & $    8.89   ^{+    1.45  
}_{-    1.46}$ & $    7.46   ^{+    2.82  }_{-    2.83}$ & $    1.11   ^{+    0.
37  }_{-    0.41 }$ \\
          $\rho_M~(10^{8}~M_{\odot}~\rm Mpc^{-3})$   & $    1.63   ^{+    0.26  
}_{-    0.27}$ & $    1.36   ^{+    0.52  }_{-    0.52}$ & $    0.26   ^{+    0.
07  }_{-    0.08 }$ \\
    \enddata
    \tablenotetext{a}{Adopted from \cite{we07}}
    \tablenotetext{b}{$z\sim0.03$ slope assumed}
  \end{deluxetable}
  \renewcommand{\baselinestretch}{1.}
}

\documentclass[aaspp4,11pt,preprint,apjfonts]{emulateapj}

\slugcomment{Accepted for publication in the Astrophysical Journal}
\shorttitle{A Red Sequence at $z\sim2.3$}
\shortauthors{Kriek et al.}

\begin{document}
  
\title{The detection of a red sequence of massive field galaxies at $z\sim2.3$ and its evolution to $z\sim0$\altaffilmark{1}}

\author{Mariska Kriek\altaffilmark{2,3}, Arjen van der
  Wel\altaffilmark{4}, Pieter G. van Dokkum\altaffilmark{5}, Marijn
  Franx\altaffilmark{3}, and Garth D. Illingworth\altaffilmark{6}}

\email{mariska@astro.princeton.edu}

\altaffiltext{1}{Based on observations obtained at the Gemini
  Observatory, which is operated by the Association of Universities for
  Research in Astronomy, Inc., under a cooperative agreement with the
  NSF on behalf of the Gemini partnership.}

\altaffiltext{2}{H.\,N. Russell Fellow, Department of Astrophysical 
  Sciences, Princeton University, Princeton, NJ 08544}

\altaffiltext{3}{Leiden Observatory, Leiden University, NL-2300 RA Leiden, 
  The Netherlands}

\altaffiltext{4}{Department of Physics and Astronomy, Johns Hopkins
  University, Baltimore, MD 21218}

\altaffiltext{5}{Department of Astronomy, Yale University, New Haven, 
  CT 06520}

\altaffiltext{6}{UCO/Lick Observatory, University of California, Santa
  Cruz, CA 95064}

\begin{abstract} 
  The existence of massive galaxies with strongly suppressed star
  formation at $z\sim2.3$, identified in a previous paper, suggests
  that a red sequence may already be in place beyond $z=2$. In order
  to test this hypothesis, we study the rest-frame $U-B$ color
  distribution of massive galaxies at $2<z<3$. The sample is drawn
  from our near-infrared spectroscopic survey for massive
  galaxies. The color distribution shows a statistically significant
  ($>3\sigma$) red sequence, which hosts $\sim60$\% of the stellar
  mass at the high-mass end. The red-sequence galaxies have little or
  no ongoing star formation, as inferred from both emission-line
  diagnostics and stellar continuum shapes. Their strong Balmer breaks
  and their location in the rest-frame $U-B$,\,$B-V$ plane indicate
  that they are in a post-starburst phase, with typical ages of
  $\sim$0.5-1.0\,Gyr. In order to study the evolution of the red
  sequence, we compare our sample with spectroscopic massive galaxy
  samples at $0.02<z<0.045$ and $0.6<z<1.0$. The rest-frame $U-B$
  color reddens by $\sim0.25$ mag from $z\sim2.3$ to the present at a
  given mass. Over the same redshift interval, the number and stellar
  mass density on the high-mass end ($>10^{11}\,M_{\odot}$) of the red
  sequence grow by factors of $\sim8$ and $\sim6$, respectively. We
  explore simple models to explain the observed evolution. Passive
  evolution models predict too strong $\Delta(U-B)$, and produce
  $z\sim0$ galaxies that are too red. More complicated models that
  include aging, galaxy transformations, and red mergers can explain
  both the number density and color evolution of the massive end of
  the red sequence between $z\sim2.3$ and the present.
\end{abstract}


\section{INTRODUCTION}

Early type galaxies with quiescent stellar populations form a
well-defined color-magnitude or color-mass relation at $z\sim0$, known
as the red sequence. They are clearly separated from blue star-forming
galaxies, which populate a different, less-tight sequence, called the
blue cloud. While the red sequence is primarily build up of massive
galaxies, blue galaxies have lower stellar
masses \citep[e.g,][]{ka03}.

The appearance and evolution of the red sequence provide a powerful
method to study the star-formation and assembly history of massive
galaxies \citep[e.g.,][]{bo92,ss92,vd98}. The red sequence exhibits a
tilt and spread which are thought to be primarily driven by
metallicity and age differences, respectively
\citep[e.g.,][]{fa73,wo94,ka97,ka99}. Both the shape and the color of
the red sequence evolve over cosmic time. Several processes are
responsible for this evolution. First, the color gradually reddens due
to aging of stellar populations. Second, the red sequence grows
through transformations of blue galaxies. These transformations change
the mix of properties of red-sequence galaxies and may cause the
observed evolution of the red sequence to deviate from the
expectations from passive evolution. Third, mergers among red-sequence
galaxies change the red-galaxy mass function and may affect the color,
slope, and scatter of the red sequence \citep[e.g.,][]{bo92}. Thus,
the evolution of the color, the shape, the number and mass density of
the red sequence sets direct constraints on the assembly and star
formation history of massive, early-type galaxies.

The evolution of the red sequence between $z\sim1$ and $z\sim0$ has
extensively been studied for this purpose. Overall, these studies find
that the color evolution is consistent with passive evolution
\citep[e.g.,][]{be04}, the mass on the red sequence doubles in this
redshift interval \citep[e.g.,][]{be04,fa07,ar07}, and the growth at
higher masses is attributed to both red mergers and galaxy
transformations \citep[e.g.,][]{bu07}. As a significant part of the
red sequence was already in place at $z\sim1$, we have to push our
studies to higher redshift to trace the onset and first build up of
the red sequence.
  
Recent high-redshift studies report the detection of the red sequence
up to $z=2$ \cite[e.g.,][]{ar07,ca08}. Moreover, developments in NIR
instrumentation have enabled the first spectroscopic confirmations of
quiescent galaxies without detected emission lines beyond $z=2$
\citep[][]{kr06a,kr06b}. In particular the
cross-dispersed mode of the Gemini Near-Infrared Spectrograph
\citep[GNIRS,][]{el06}, with a wavelength coverage of 1-2.5 $\mu$m
allows systematic studies of massive galaxies at $z\sim2.3$. Using
this instrument we have completed a NIR spectroscopic study of 36
$K$-selected galaxies at $2\lesssim z \lesssim 3$. In this paper we
use this survey to study the onset and color evolution of the red
sequence. The spectroscopic redshifts in combination with the accurate
continuum shapes as provided by the NIR spectra, allow for the first
time accurate rest-frame color determinations of quiescent, massive
galaxies beyond $z=2$.

Throughout the paper we assume a $\Lambda$CDM cosmology with
$\Omega_{\rm m}=0.3$, $\Omega_{\Lambda}=0.7$, and $H_{\rm 0}=70$~km
s$^{-1}$ Mpc$^{-1}$, and a \cite{sa55} initial mass function (IMF)
between 0.1 and 100 $M_{\odot}$. All broadband magnitudes are given in
the Vega-based photometric system.

\section{DATA}\label{sec:data}

The data used in this work are extracted from our NIR spectroscopic
survey for massive galaxies \citep[][]{kr08}. The full sample consists
of 36 $K$-bright galaxies observed with GNIRS in cross-dispersed mode
(1.0--2.5 $\mu$m), between 2004 September and 2007 March (programs:
GS-2004B-Q-38, GS-2005A-Q-20, GS-2005B-C-12, GS-2006A-C-6,
GS-2006B-C-5 and GS-2007A-C-9). The galaxies were originally selected
from the multi-wavelength survey by Yale-Chile
\citep[MUSYC,][]{ga06,qu07}, which provides us with accurate
optical-to-NIR ($UBVRIzJHK$) photometry. 

In \cite{kr08} we show that distribution of the rest-frame $U-V$ and
observed $R-K$ colors of our spectroscopic sample are representative
of a mass limited sample at $2<z<3$. However, we do note that we are
biased towards galaxies with brighter $K$-band magnitudes. Although
this mainly reflects the relatively lower redshifts of these galaxies,
as explained in detail in Kriek et al. 2008, we might be missing
galaxies with higher mass-to-light ratios ($M/L$). Further details
about sample completeness, observations, reduction and extraction of
the spectra can also be found in \cite{kr08}.

For this work we use the 28 galaxies within the range $2<z_{\rm
spec}<3$. Stellar masses and other population properties are derived
by stellar population modeling as described in detail in
\cite{kr06a,kr08}, and given in Table 2 in \cite{kr08}. In summary, we
fit the spectra together with the broadband optical photometry by
\cite{bc03} stellar population models, assuming an exponentially
declining star formation history, solar metallicity, the \cite{ca00}
reddening law and the \cite{sa55} IMF between 0.1 and 100
$M_{\odot}$. We allow a grid of 41 values for $A_V$ between 0 and 4
mag, 31 values for the characteristic star-forming timescale ($\tau$)
between 10 Myr and 10 Gyr, and 24 values for age (not exceeding the
age of the universe). Uncertainties on the stellar population
properties are derived using 200 Monte Carlo simulations as described
in \cite{kr06a,kr08}. We leave redshift as a free parameter for
galaxies without emission lines. We tested the continuum redshifts
using the emission-line galaxies in our sample, and found an
uncertainty in $\Delta z/(1+z)$ of less than 0.019 \citep{kr08}.

Rest-frame $U-B$\footnote{Throughout this paper we use the Buser U, B3
and V filters.} colors are also determined from the best-fit stellar
population models. In the same fashion as for the stellar population
properties, confidence levels are derived from Monte Carlo
simulations. Hence, the confidence intervals on the rest-frame colors
include the uncertainties on the continuum redshifts for galaxies
without emission lines. The colors are not measured directly from the
spectra, as for several galaxies the $U$-band is not covered
completely by the NIR spectrum, or the $B$-band falls partly in
between the $J$ and $H$ atmospheric windows. As our grid allows almost
30~000 different synthetic spectra, we do not expect the colors to
converge to certain best-fit templates. Nevertheless, in order to
examine whether using best fits may introduce systematics in the
derived rest-frame colors, we directly measure the colors from the NIR
spectra in combination with the optical broadband photometry. We find
no systematic offset between the direct colors and those derived from
the best fits.

The unique aspects of this data set are the accurate spectroscopic
redshifts, rest-frame colors and masses. Although broadband photometric
studies provide much larger galaxy samples, they lack the accuracy
needed for this study.

\figa

\section{A RED SEQUENCE AT $\lowercase{z}\sim2.3$}\label{sec:rs_hz}

\subsection{The Detection of the Red Sequence at $z\sim2.3$}\label{sec:det_rs}

Figures~\ref{fig:ub_mass}a and b present rest-frame $U-B$ color versus
stellar mass for the $2<z<3$ massive galaxy sample. The colors are
corrected for redshift differences within the sample, as will be
explained in \S~\ref{sec:rs_col}. These correction are very small
($\sim$0.001), and barely change the appearance of this plot. The
distribution of galaxies in Figures~\ref{fig:ub_mass}a and b is
striking, as there are many galaxies with similar, red colors. In
order to test whether a red sequence was already in place at this
early epoch, we examine the rest-frame color distribution. But first
we correct the colors for the tilt of the red sequence, by assuming
the slope of $z=0$ red sequence \citep{we07}. The applied correction
has the form
\begin{equation}
\label{eq:ub_m}
(U-B)_{M} = (U-B) - 0.08 \ ( {\rm log} M / M_{\odot} - 11.3 )
\end{equation}
The residuals are shown by the black histograms in
Figures~\ref{fig:ub_mass}c and d. The distribution exhibits a
conspicuous peak at $(U-B) \sim 0.25$ mag. We test the significance of
the red sequence by calculating the probability to obtain this peak
when assuming a flat distribution in rest-frame $U-B$ color. If the
true distribution is uniform the probability of finding nine or more
galaxies in any one of the three red bins and finding 16 or more
galaxies in total with $U-B>0.1$ is less than 0.001. Thus the change
of finding the detected red sequence by change is less than
0.1\%. This implies that the detection of the red sequence is
significant at the $>3\sigma$ level. This all strongly suggests that a
red sequence of massive field galaxies was most likely already in
place at $z\sim2.3$.
 
We determine the location of the peak of the red sequence by fitting a
Gaussian to the color distribution. We average over many binning
positions to obtain a distribution that is not affected by the
particular choice of binning. In order to avoid including blue-cloud
galaxies, we restrict the fitting region to all galaxies with
$(U-B)_{M} > (U-B)_{\rm peak}-0.1$. Hence, this procedures requires a
few iterations. The cut-off value of 0.1 mag is chosen as it is twice
the scatter in rest-frame $U-B$ color of the local red sequence (see
\S~\ref{sec:rs_col}).  The best fit is indicated by the solid curve in
Figures~\ref{fig:ub_mass}c and d. The peak of the distribution is
shown by the solid orange line in Figures~\ref{fig:ub_mass}a and
b. All galaxies above the dotted line ($[U-B]_{M} > [U-B]_{\rm
  peak}-0.1$) are defined as red-sequence galaxies from hereon.

Our result may seem in disagreement with previous studies, some of
which indicate that the red sequence disappears beyond $z=1.5$
\citep[e.g.,][]{ci07}. However, finding a red sequence, as is well
known, requires very accurate rest-frame color determinations. The use
of a galaxy sample with spectroscopic redshifts and stellar continuum
shapes -- {\em and not just photometric information} -- enables us to
detect a significant red sequence beyond $z=2$, in contrast to previous
studies. Broadband photometry in combination with photometric
redshifts with errors of $\Delta z/(1+z)\sim 0.07$ gives random errors
of 0.1 mag in rest-frame $U-B$, and systematic errors may play an even
larger role. Furthermore, the typical uncertainties on stellar mass
and absolute magnitude are a factor of $\sim2$ and $\sim0.4$ mag,
respectively \citep{kr08}. Thus, the uncertainties on the location of
the individual red-sequence galaxies are larger than the width of the
intrinsic red sequence. This implies that photometric studies with
errors of $\sim0.07$ in $\Delta z/(1+z)$ are not able to recover a red
sequence. Studies that lack spectroscopic redshifts in the relevant
redshift range
\citep[e.g.,][]{ci07} to calibrate their photometric redshifts, most
likely have even larger errors in $\Delta z/(1+z)$. 

Whereas we do detect a significant red sequence, we find no bimodality
in the galaxy distribution. We cannot reliably comment on bimodality
since larger samples over a larger mass range are needed to assess
whether the blue galaxies are distinct from the red ones at this
epoch. Nonetheless, recent work by \cite{ca08}, based on a
spectroscopic study over a larger stellar mass range, shows that the
galaxy bimodality exist at least out to a redshift of $z=2$. We note
that the study by \cite{ca08} is based on spectroscopic redshifts,
reinforcing our hypothesis that the current result can only be
obtained with redshifts more accurate than the standard photometric
redshifts.

Finally, we stress that the detection of a red sequence is independent
on whether the sample is fully representative of a mass-limited
sample. In this context it is interesting to note that the red
sequence has originally been discovered and mainly studied in
magnitude- and not mass-limited samples.

\subsection{Properties of Red-Sequence Galaxies at
  $z\sim2.3$}\label{sec:pop_rs}

In the previous section we showed that a red sequence was already in
place at $z\sim2.3$. The well-defined shape of the red sequence could
be a consequence of the converging colors of evolved
galaxies. However, our sample is small, and dusty starburst galaxies,
known to be highly abundant at these
redshifts \cite[e.g.,][]{we06,pa06}, may contribute to, or even
dominate the red sequence at these early epochs. In order to test
whether these red-sequence galaxies indeed host quiescent stellar
population, we examine the star formation properties using several
diagnostics.

In Figure~\ref{fig:ub_mass}a the galaxies are coded following their
best-fit specific SFR derived from modeling their stellar continua
\citep[see \S~\ref{sec:data} and ][]{kr08}. The corresponding color
distributions are presented in Figure~\ref{fig:ub_mass}c. Most
red-sequence galaxies are best-fit by specific SFRs less than
0.05~Gyr$^{-1}$, and three have specific SFRs between 0.05 and
0.5~Gyr$^{-1}$. The uncertainties on the specific SFRs are about a
factor of $\sim3$ on average \citep[][]{kr08}. Nevertheless, the large
fraction of galaxies with low specific SFRs suggests that red sequence
is not dominated by dusty starbursts.

A combination of two rest-frame colors, such that one isolates the
optical break, and the other color measures the slope of the spectrum
redwards of the optical break \citep[e.g.,][]{fo04,la05,wu07}, may
also be used to discriminate between dusty star-forming galaxies and
quiescent stellar populations. In Figure~\ref{fig:ub_bv} we show
rest-frame $U-B$ versus $B-V$ for all galaxies at $2<z<3$. Color
evolution tracks of \cite{bc03} models show that quiescent stellar
populations have a different locus than dusty starbursts. As expected,
the red-sequence galaxies are closer to the simple stellar population
(SSP) model tracks. The remaining galaxies can roughly be divided in
those that have colors more comparable to constant star forming (CSF)
models with dust, and galaxies that will probably soon join the red
sequence.

\figb
\figc

These results are supported by independent emission line diagnostics,
presented in Figure~\ref{fig:ub_mass}b and d. We divide the sample
according to whether emission lines are detected in the rest-frame
optical spectra. Subsequently, the emission-line galaxies are sorted
for the dominant origin of their line emission: using primarily
emission-line ratios we discriminate between active galactic nuclei
(AGNs) and H\,{\sc ii} regions \citep[][]{kr07}. For two galaxies we
have no information on the line emission as the lines are expected at
wavelengths with low atmospheric transmission. 13 out of 15
red-sequence galaxies have no detected emission lines, or the line
emission is dominated by AGNs.

Thus, both the stellar continua and the emission line diagnostics
suggest that the red sequence at $z\sim2.3$ is dominated by galaxies
with quiescent stellar populations. In particular, 7 out of 9
previously identified galaxies with strongly suppressed star formation
presented in \cite{kr06b} fall on this red sequence. The rest-frame
$U-B$ colors of the two remaining galaxies with strongly suppressed
star formation are just below the cut-off value, and their locus in
Figure~\ref{fig:ub_bv} is near the SSP track. Thus, they will most
likely soon join the red sequence.

The rest-frame $B-V$ colors in combination with $U-B$ indicate that
the red-sequence galaxies at $z\sim2.3$ are likely in a post-starburst
phase (Figure~\ref{fig:ub_bv}). This is further illustrated in
Figure~\ref{fig:stack}, in which we show the stacked low-resolution
spectrum of all red-sequence galaxies. For comparison we show a 2 Gyr
SSP model with a prominent 4000 \AA\ break as well. In contrast to
such old stellar populations, the optical break for the $z\sim2.3$
red-sequence galaxies is clearly dominated by the Balmer
break. Overall, our findings may imply that the massive end of the red
sequence is just starting to build up at $z\sim2.3$, and was likely
not yet in place beyond $z\sim3$. \cite{ko07} drew the same conclusion
by studying the stellar populations in protoclusters at $2\lesssim z
\lesssim3$. Our work is also consistent with the study by \cite{br07}
who found that in contrast to $z\sim2.4$, red galaxies at $z\sim3.7$
have significant UV emission and are thus still actively forming
stars. We note, however, that as explained in \S\,2, we might be
missing galaxies with higher $M/L$, as these are relatively faint in
$K$.

Furthermore, Figure~\ref{fig:ub_bv} provides us with a clear
illustration of the different processes that may be responsible for
the spread and the tilt of the red sequence at $z\sim2.3$. $U-B$ and
$B-V$ show a positive correlation for the red-sequence galaxies. The
typical 1$\sigma$ confidence contour shows that random errors can not
fully account for the spread, and other effects are likely to play a
role. The two SSP tracks, indicated by the dashed lines, show that
both age and metallicity differences may be responsible for the spread
and the tilt. Also the redshift spread of the red sequence galaxies,
corresponding to 0.5 Gyr, will induce scatter in colors. Finally,
reddening by dust moves a galaxy in a similar direction as aging and
metallicity, and may also contribute to the spread and the tilt of the
red sequence. The importance of the different processes can not be
addressed with the current data, and independent dust, age, and
metallicity constraints are required to break the degeneracies.

Finally, we stress that, although the star formation activity in the
red-sequence galaxies appears low, the galaxies may still be reddened
by fair amounts of dust. Best-fit stellar population models indicate
an average dust content of $A_V=0.8$\,mag (using the Calzetti
reddening law). However, due to degeneracies between age and dust,
$A_V$ is poorly constrained with typical uncertainties of 0.5\,mag
\citep[][]{kr08}, and dust-free models provide almost
equally good fits to the spectra \citep[][]{kr06b}. Metallicity ($Z$)
further complicates this degeneracy, and as it is fixed to $Z_{\odot}$
during fitting, the lack of appropriate metallicities may have been
compensated by adjusting age or $A_V$. Independent indicators, such as
mid-infrared (MIR) imaging or Balmer decrements are needed to better
constrain the dust content in these red-sequence galaxies. MIR imaging
also reveals whether obscured star-forming regions may have been
missed.  In this context it is interesting to note that \cite{re06a}
and \cite{pa06} show that about half of the galaxies at the high
mass-end of the galaxy distribution at $z\sim2.3$ are not detected at
24$\micron$. Moreover, \cite{re06a} find that the H$\alpha$ luminosity
for $z\sim2$ galaxies tracks the bolometric luminosity very
well. Finally, large amounts of dust in all red-sequence galaxies
studied in this work seems unlikely, because it would then be hard to
explain the narrow peak in the observed color distribution.

\section{MEASURING THE EVOLUTION OF THE RED SEQUENCE}
\label{sec:meas_rs}

In order to measure the rest-frame $U-B$ evolution of the red
sequence, we compare our $z\sim2.3$ results with those of lower
redshift samples in this section. We use two spectroscopic massive
galaxy samples ($>10^{11} M_{\odot}$) at $0.02<z<0.045$ and
$0.6<z<1.0$, extracted from the Sloan Digital Sky Survey
\citep[SDSS,][]{yo00}, Data Release 5 \citep[DR5;][]{ad07} and the
Great Observatories Origins Deep Survey \citep[GOODS;][]{gi04},
respectively.

\subsection{Spectroscopic Samples at Lower Redshifts}\label{sec:dat_lz}

For our lowest redshift sample we use a complete, mass-selected,
volume-limited sample of galaxies at redshifts $0.02<z<0.045$,
extracted from the SDSS. See \cite{we07} for more details about the
extraction and completeness of the sample.  The SDSS $u-g$ and $g-r$
colors are used to derive rest-frame $U-B$ colors\footnote{The AB to
Vega zeropoint conversion for the $U$, $B$, and $V$ bands used in this
paper are slightly different from those adopted in \cite{we07}. The
conversions used in this paper are the same as used by
\cite{bc03}.}. $M/L$ are derived from the $g-r$ colors (corrected for
galactic extinction and redshift) using the relation by
\cite{be03}. The inferred stellar masses are increased by 0.15 dex
to account for differences in the IMF. The final massive ($>10^{11}
M_{\odot}$) galaxy sample consists of 903 galaxies.

For the intermediate redshift sample at $0.6<z<1.0$, we use a
mass-selected, volume-limited galaxy sample, constructed from
GOODS-south. A detailed description of the extraction of the sample
and completeness can be found in \cite{we07}. Rest-frame $U-B$ and
$B-V$ colors for the $0.6<z<1.0$ sample are derived from the F606W,
I775W and F850LP ACS photometry. Subsequently, $B-V$ provides us
with stellar masses, using the empirical relations by
\citet{be03}. Again, the stellar masses are increased by 0.15 dex. Of
this sample, 137 galaxies have masses $>10^{11}
M_{\odot}$. Spectroscopic redshifts are known for 70\% of this sample
\citep[][]{lf04,mi05,we05,va06}. As this spectroscopic sample is not
fully representative of the total sample, we include the galaxies
with photometric redshifts \citep{wu08} when appropriate.

\figd

\subsection{The Color Evolution of the Red Sequence}\label{sec:rs_col}

For an accurate measurement of the $U-B$ color evolution, it is
crucial to derive the color of the red sequence in a similar fashion
as for our $z\sim2.3$ sample. In the top panels of
Figure~\ref{fig:ub_hist} we show rest-frame $U-B$ versus stellar mass
for all three massive galaxies samples. The colors are corrected for
redshift differences within each subsample, in a self-consistent way,
by using the evolution of the red-sequence color with time derived
below. Next, we subtract equation~(\ref{eq:ub_m}) from the data and
determine the peak of the color distribution in the same way as was
done for the $z\sim2.34$ sample. The extracted color distribution
along the $z\sim0.0$ slope is presented in the lower panels of
Figure~\ref{fig:ub_hist}. Galaxies with colors $(U-B)_{M, z} >
(U-B)_{\rm peak}-0.1$ are defined as belonging to the red
sequence. The peak locations are indicated by solid lines in
Figure~\ref{fig:ub_hist}. For the $0.6<z<1.0$ galaxies we exclude the
30\% without spectroscopic redshifts when deriving the peak of the red
sequence, as these galaxies have less accurate rest-frame colors.

Random errors on the location of the red sequence are determined using
bootstrapping, and are represented by the dotted lines in the lower
panels of Figure~\ref{fig:ub_hist}. The use of different samples and
different method to derive rest-frame colors and stellar masses may
introduce additional systematic errors. In order to correctly
interpret the observed evolution, we examine the following
effects. \smallskip

--~{\it Aperture differences} \ Elliptical galaxies exhibit color
gradients \citep[e.g.,][]{fi90}. Thus, differences in the apertures
that were used to measure rest-frame $U-B$ may introduce systematics
in the color evolution. For both the $z\sim0.03$ and $z\sim0.73$
sample the rest-frame $U-B$ colors are derived for an aperture which
is about equivalent to the half-light radius. For the $z\sim2.3$
sample we use the NIR spectra to measure rest-frame $U-B$. The
aperture sizes are rectangular and depend on the slitwidth (of
0\farcs675), the extraction aperture, and extraction method. We use a
weighted (with S/N) extraction and include spatial elements that have
a flux larger than $0.25$ times the maximum
flux \citep[][]{kr08}. Thus, the total aperture size is dependent on
the light distribution of the galaxy. The average effective aperture
of the $z\sim2.34$ sample is comparable to a circular aperture of
$\sim8$\,kpc.

In order to quantify possible systematics in rest-frame $U-B$
determinations of red-sequence galaxies between the $z<1$ and
$z\sim2.3$ samples, we measure rest-frame $U-B$ for all galaxies in
the $z\sim0.73$ sample using an aperture of 1\arcsec ($\sim$7.6 kpc at
$z\sim0.73$) instead of 0\farcs5. This aperture size is comparable to
that used for the $z\sim2.3$ sample. For the increased aperture the
$U-B$ color of the $z\sim0.73$ red sequence is 0.01 mag bluer. In
order to correct for this effect, we have reduced the colors of the
$0.02<z<0.045$ and $0.6<z<1.0$ galaxies by 0.01 mag.\smallskip

--~{\it Zero points} \ Uncertainties in zero points directly result in
systematic uncertainties in the derived rest-frame colors.  The
uncertainty on individual zero points is $<0.01$ mag for both the
$0.02<z<0.045$ \footnote{http://www.sdss.org/} and $0.6<z<1.0$
\citep{si05} samples. Thus, the resulting uncertainty on $U-B$ is
$<0.02$ mag. For the $z\sim2.3$ sample we use AV\,0-type stars for
calibrating the spectra. This results in an uncertainty on rest-frame
$U-B$ of $\sim0.03$ mag. For this sample the broadband photometry
provides us with an independent check on the zero points. In
\cite{kr08} we find a systematic offset of $\sim0.01$ mag between
rest-frame $U-V$ derived from the photometry (using $z_{\rm spec}$)
and the NIR spectra. This shows that the zero points probably
introduce no large systematics for the $z\sim2.3$ sample.\smallskip

--~{\it Rest-frame color determinations} \ The determination of
rest-frame colors may also result in systematic errors. For the low
and intermediate redshift samples we use the photometry in combination
with spectroscopic redshifts. This method may result in systematics of
$\sim0.01$ mag and $\sim0.03$ mag for the $0.02<z<0.045$ and
$0.6<z<1.0$ samples, respectively. For the $z\sim2.3$ sample we have
higher resolution spectral shapes, provided by the NIR spectra. In the
previous paragraph and in \S~\ref{sec:data} we discussed two tests to
assess rest-frame color determinations, and neither of them exhibit
significant systematics ($\sim0.01$ mag).\smallskip

--~{\it Stellar mass determinations} \ Although the stellar masses are
all determined for the same IMF, the different methods used for the
$z<1$ and $z\sim2.3$ samples may have introduced systematic errors. We
tested this by determining the stellar masses for the $z\sim2.3$
galaxies using the same method as for the lower redshift samples. This
yielded stellar masses which are typically 0.13 dex less than the
best-fit stellar masses. This may be due to younger and dustier
stellar populations of the $z\sim2.3$ massive galaxies. As the slope
of the red sequence is 0.08 mag dex$^{-1}$, this may result in a
systematic error of $\sim$0.01 mag on the observed color
evolution. Furthermore, this effect may also alter the cut-off value
of $10^{11} M_{\odot}$, and consequently the selection of massive
red-sequence galaxies. Increasing the mass cut-off value by 0.13 dex
would not change the peak location of the red sequence. However, the
number and mass fractions of massive galaxies on the red-sequence
would increase by $\sim$10\%.

\smallskip

--~{\it Completeness} \ The $z\sim0.03$ sample is volume limited, and
thus completeness effects are not expected to play a role. For the
$z\sim0.73$ sample we use just the spectroscopic subsample to estimate
the red-sequence color, as photometric redshifts results in less
accurate rest-frame colors. However, including all galaxies, would
give a rest-frame $U-B$ color which is 0.01 mag less. Thus,
completeness effects may result in a systematic error of $\sim0.01$
for the $z\sim0.73$ sample.

\fige

The $z\sim2.34$ sample is much smaller and not volume limited. In
\cite{kr08} we find that the distributions of rest-frame
$U-V$ color, observed $R-K$ and $J-K$ color and redshift for the
spectroscopic sample at $2<z<3$ are similar as for a photometric mass-
and volume-limited sample at the same redshift
interval. Unfortunately, we can only compare photometric properties in
order to investigate whether the subsample is representative, and thus
systematics in photometric studies may jeopardize the real
completeness. For example, in \cite{kr08} we identified systematics
between photometric redshift and SED type, such that dusty, young
galaxies were generally placed at too high redshifts. As these dusty
galaxies scattered to lower redshift, they are not included in the
sample used in this work. Dusty galaxies with $2<z_{\rm spec}<3$ may
not be properly represented in the sample, as they were initially
placed at too high redshift. Although this possible incompleteness may
not alter our findings of the color of the red sequence, it should be
kept in mind that the sample may not be complete and representative
of the total population of $2<z<3$ galaxies. We estimate that
completeness effects may result in a systematic error of $\sim0.03$
mag on the rest-frame $U-B$ color of the $z\sim2.3$ red sequence.

The various effects discussed above result in total systematic
uncertainties of 0.03, 0.04 and 0.05 mag for the $z\sim0.0$,
$z\sim0.7$ and $z\sim2.3$ red-sequence colors respectively. We assume
that the random and systematic errors are independent and can be added
in quadrature.

The color evolution of massive galaxies is shown in
Figure~\ref{fig:ub_t}. The small filled dots show individual
red-sequence galaxies, and the large filled symbols with errorbars
show the peak locations for galaxies on the red sequence. The dotted
line is a simple linear fit to the large symbols, of the form:
\begin{equation} 
\label{eq:ub_t} 
(U-B)_{z} = 0.19 + 0.33~t
\end{equation}
with $t$ the fractional age of the universe. The fit was used to apply
differential color corrections to account for redshift differences
within each sample (see above), and provides a remarkably good fit.

Table~\ref{tab:rs} lists the rest-frame $U-B$ color and the width of
the red sequence for the three samples. The color of the red sequence
evolves by $\sim$0.25 mag between between $z\sim2.3$ and
$z\sim0.0$. In \S~\ref{sec:mod_rs} we attempt to explain this
evolution using simple models.

\taba

\subsection{Evolution of the Mass and Number
  Density}\label{sec:rs_dens}

In addition to $\Delta(U-B)$, the evolution of the number and mass
density of red-sequence galaxies places constraints on the build up of
the red sequence. The densities for the $z<1$ samples follow directly
from the used samples, as both are volume limited. To obtain the
relative fractions at $z\sim0.73$ we use the full $0.6<z<1.0$ sample,
including galaxies with photometric redshifts, in combination with the
red-sequence location as derived from the spectroscopic sample.

The galaxies at $2<z<3$ do not form a complete sample. We use the
MUSYC deep survey \citep[1030, 1256 and HDF-South,][]{qu07} to
estimate the total number of massive galaxies ($>10^{11} M_{\odot}$)
at $2<z<3$ (see \S~\ref{sec:data}). Note that most galaxies of our
spectroscopic sample are extracted from this survey. We correct the
obtained density for systematics in photometric redshifts as derived
in \cite{kr08}. We find a total number density of massive galaxies of
$\rho~(M>10^{11}M_{\odot}) = 2.0 \times 10^{-4} \rm ~Mpc^{-3}$. This
value is consistent with the values found by \cite{vd06} of
$2.2^{+0.6}_{-0.6} \times 10^{-4} \rm ~Mpc^{-3}$ and by \cite{dr05} of
$\sim1.8 \times 10^{-4} \rm ~Mpc^{-3}$ for the same mass-cut and
IMF. Subsequently we use the mass and number fractions, as estimated
from the spectroscopic sample to derive the mass and number densities
of red-sequence galaxies. We find a number density of massive
red-sequence galaxies at $z\sim2.3$ of $1.1^{+0.4}_{-0.4}
\times 10^{-4} \rm ~Mpc^{-3}$.

All number and mass densities are given in Table~\ref{tab:rs}. The
uncertainties on the fractions are determined by bootstrapping (see
\S~\ref{sec:rs_col}). For the number and mass densities, the
uncertainties include cosmic variance, random errors, uncertainties on
the fractions, and uncertainties introduced by different stellar mass
estimators. The {\em fraction} of the total massive galaxy population
that is on the red sequence has evolved by only $\sim20$\% from
$z\sim2.3$ to the present. Similarly, the fraction of the total
stellar mass in galaxies with $M>10^{11} M_{\odot}$ that is on the
red sequence has evolved by $\sim15$\%. However, the total number and
total mass of red sequence galaxies has increased by factors of $\sim8$
and $\sim6$ respectively over this time interval.

\subsection{Comparison with Other Studies}

In this section we discuss our results in the context of other studies
that investigated the evolution of the red sequence. However, when
comparing our findings to these studies, it is important to keep in
mind the difference in sample selection and red-sequence
definition. For example,
\cite{sh05} and \cite{re06b} find no red-sequence galaxies among
their $z\sim2$ Lyman Break Galaxies \citep[LBGs][]{st96a,st96b}, as
the galaxies are selected by their bright optical flux, and thus they
only target star-forming galaxies. Only 20\% of a mass-limited sample
would be identified by LBG selection, and thus this selection
technique is not suitable for obtaining unbiased mass-limited samples
\citep{vd06}.

Using a mass-limited sample with a wide variety of data over a total
field of 1.53 $\rm deg^2$, \cite{co07} find that the red-sequence
fraction of galaxies with $10^{11} M_{\odot} < M_* < 10^{11.5}
M_{\odot}$ increases from 0.65 to 0.88 between $z\sim1.4$ and
$z\sim0.4$. Direct comparison with our work is complicated by the
different mass range, a different adopted IMF
\citep{ch03}, and different color criteria to identify red-sequence galaxies. 
Nevertheless, the slight decrease of this fraction may be consistent
with the findings presented in this work.

Our study is also broadly consistent with the results of the K20
survey. \cite{fon04} find that $\sim$30-40\% of the present day
stellar mass in objects with $5
\times 10^{10} M_{\odot} < M_* < 5 \times 10^{11} M_{\odot}$ appears
to be in place at $z\sim2$. We find that for $M_* > 1 \times 10^{11}
M_{\odot}$ 20\% of the mass was already in place at a slightly higher
redshift of $z\sim2.3$. Given the large uncertainties of both studies,
the different mass cuts, and the different targeted redshift range,
these results are not in conflict with each other. Their relative
fraction of red to blue sequence galaxies is difficult to compare to
ours, as K20 has spectroscopic redshifts and classifications for 43\%
of the galaxies at $1.3<z<2.0$. 21\% are early types and 22\% are late
types. So for more than half of the sample the nature of the galaxy is
unknown. As late types are easier to spectroscopically confirm
(especially with optical spectroscopy) than early types, it is not
unlikely that the early types span the majority at this redshift
range. Thus, their result is not inconsistent with the fractions we
find for our spectroscopic sample at z$\sim$2.3.

\cite{ci07} find that less than half of the galaxies exceed
their red-sequence cut. However, these authors use a $K$-selected,
instead of a mass-limited sample. $K$-selection gives higher fractions
of blue galaxies, as such galaxies are brighter in $K$ than red
galaxies at the same stellar mass. Also, since their color
distribution shows no obvious peak, their definition of the
red-sequence galaxies cannot be directly compared to ours.

Finally, \cite{ca08} identify color bimodality out to z=2 among the
galaxies in the GMASS survey. Their sample has a spectroscopic
completeness of 50\%, with 190 spectroscopic redshifts beyond
z=1.4. The authors show that early-type galaxies make up 50\% of the
population at the high mass end at $z=2$. When imposing the same color
and mass cut as used in this paper, we derive that red-sequence
galaxies would dominate. Furthermore, in agreement with our study,
they find that the rest-frame $U-B$ color evolves by only $\sim0.2$
mag between $z\sim2.5$ and $z\sim0.5$.

In summary, it is difficult to directly compare our results with other
studies, as few in the targeted redshift range use mass limited
samples. Even more importantly, as already discussed in \S~3.1, the
lack of spectroscopic information prevents accurate rest-frame color
determinations of red galaxies for current $z>2$
studies. Nevertheless, our study seems broadly consistent with other
work.

\figf
\figg

\section{MODELING THE EVOLUTION OF THE RED SEQUENCE}
\label{sec:mod_rs}

In the previous section we derived the evolution of rest-frame $U-B$
color and the number density for the massive end ($>10^{11}
M_{\odot}$) of the red sequence between $z\sim2.3$ and the present. In
this section we attempt to explain this evolution using simple models.

\subsection{Aging of Stellar Populations}\label{sec:age}

The red-sequence galaxies at $z\sim2.3$ have no or very little ongoing
star formation and thus are expected to evolve passively over
time. Therefore, simple aging of stellar populations seems the most
straightforward explanation for the observed color evolution. In
Figure~\ref{fig:dub_t}a we compare the measured $U-B$ color evolution
to single burst, solar metallicity models by both \cite{bc03} and
\cite{ma05}. In \S~\ref{sec:rs_hz} we determine that the $z\sim2.3$
galaxies are in a post-starburst phase, with typical ages of $\sim$0.5
and $\sim$1 Gyr for dust and dust-free models, respectively
\citep{kr06b,kr08}. Therefore, we assume a formation redshift of the
stars of $z_{\rm form}=3.6$. For comparison we also give a subsolar
model ($0.4Z_{\odot}$) and a model with infinite formation redshift,
both from the \cite{bc03} library.

Figure~\ref{fig:dub_t}a shows that passive stellar population models
with $z_{\rm form}=3.6$ and solar metallicity predict a too strong
color evolution, and produce galaxies at $z\sim0$ that are too red. Only
sub-solar models can explain the observed evolution at $2\times 10^{11}
M_{\odot}$. However, this is in contradiction with studies of local
ellipticals with similar stellar masses, which imply solar to
supersolar metallicities \citep[e.g.,][]{wo92}.

Perhaps more importantly, passive evolution models do not predict any
evolution in the density of galaxies on the red sequence, in clear
conflict with the observations (see solid curve in
Fig.~\ref{fig:dnd_t}). Therefore, the evolution of the red sequence is
more complicated than just aging. In the following sections we discuss
different processes that may have depressed the color evolution.

\subsection{Growth of the Red Sequence by Transformations}\label{sec:prog}

Not all galaxies stop forming stars at high redshift, and the red
sequence will be constantly supplemented by newly quenched
galaxies. Galaxies that move to the red sequence at later times will
be bluer than the galaxies that already reside on the red sequence,
and subsequently will flatten the total color evolution. This effect
is known as the progenitor bias \citep{df01}. Qualitatively, such
models can reduce the apparent color evolution while there is rapid
density evolution and a constant influx of relatively young galaxies
to the red sequence.

We apply a simple model in order to test whether including progenitor
bias can explain the observed evolution. We assume a formation period
of 100 Myr around $z_{\rm form}$ in which all galaxies are formed. The
individual galaxies are quenched at $t_{\rm stop}$, following the
probability distribution of $\tau_{\rm stop}$:
\begin{equation}
  \label{eq:t_stop} 
  P(t_{\rm stop}) \propto {\rm exp}(-t/\tau_{\rm stop})   
\end{equation} 
We use \cite{bc03} models to construct the composite evolution from
the color evolution of individual galaxies. We simulate 10~000
galaxies and determine the red-sequence color using the same method as
used for the data (see
\S~\ref{sec:rs_col}). In contrast to \cite{df01}, we assume a constant
SFR between $t_{\rm start}$ and $t_{\rm stop}$.

In Figure~\ref{fig:dub_t}b we examine whether the supply of newly
quenched galaxies can explain the observed $U-B$ evolution. The dotted
curve represents the evolution for an exponentially declining
quenching model ($\tau_{\rm stop}=\,1\,\rm Gyr$). The formation time
has been set such that, consistent with the observations, the
$z\sim2.3$ red-sequence galaxies have stopped forming stars for
$\sim$1~Gyr ($z_{\rm form}~=~4.7$). This model reduces $\Delta (U-B)$
compared to just aging (solid line), but matches neither the observed
color nor the number density evolution (see Fig.~\ref{fig:dnd_t}).

One way to further decline the color evolution and better match the
observed number density evolution, is increasing $\tau_{\rm stop}$ to
infinity. However, a more realistic model would be to vary $t_{\rm
  start}$ in addition to $t_{\rm stop}$. We define the probability
distribution of the formation time of the galaxies as
\begin{equation}
  \label{eq:t_form} 
  P(t_{\rm start}) \propto {\rm exp}(-t/\tau_{\rm start})   
\end{equation}
The dashed curve in Figure~\ref{fig:dub_t}b represents the model with
exponentially declining formation and quenching rates ($\tau_{\rm
start}~=~1~\rm Gyr$ and $\tau_{\rm stop}~=~1~\rm Gyr)$.  Also for this
model we assume a minimum formation redshift such that on average
red-sequence galaxies at $z\sim2.3$ have quenched their star formation
1 Gyr ago ($z_{\rm form}~=~12$).  This model provides a better fit to
$\Delta (U-B)$ than the previous two models, and matches the evolution
of the number density of massive red-sequence galaxies (see
Fig.~\ref{fig:dnd_t}). Nonetheless, the predicted color evolution for
$2\times 10^{11} M_{\odot}$ red-sequence galaxies is still too slow,
and other processes may be needed to match the color evolution. We
note, however, that the stellar population models by \cite{ma05} would
have provided a slightly better fit to the observed color evolution.

\subsection{Red Mergers}\label{sec:merge}

Mergers on the red sequence may also alter the color evolution
\citep[e.g.,][]{bo98}. Assuming that no star-formation is triggered or
other major processes take place, merging of two red sequence galaxies
increases the stellar mass, but leave the color unchanged. Thus red
galaxy mergers shift the red sequence to higher masses, and
consequently reduce the color evolution when measured at fixed mass.

The dashed-dotted curve in Figure~\ref{fig:dub_t}b represents the
color evolution when including red mergers, in addition to aging and a
varying $t_{\rm start}$ and $t_{\rm stop}$. We assume a constant
merger rate, normalized such that red-sequence galaxies experience one
equal mass merger between $z\sim1$ and the
present. Figure~\ref{fig:dub_t}b shows that the combined model
provides a reasonable fit to the observed evolution. If we ignore
galaxy transformations, we need 5-7 major mergers between $z\sim2.3$
and the present to provide a good fit to the observed color evolution.

Observational evidence for red mergers
\citep[e.g.,][]{vd05,tr05,be06} validates this
explanation. However, the corresponding decline in the rest-frame
$U-B$ color evolution in this study might be overestimated. First, our
assumed merger rates between $z\sim1$ and $\sim0$ may be too high, as
accurate observational constraints are still lacking. Furthermore, not
all major mergers are equal mass mergers, and for example a 3:1 merger
($\Delta[U-B]$=0.020 mag) has less impact on the evolution than a
1:1 merger ($\Delta[U-B]$=0.024 mag). 

The effects of red mergers on the number density evolution are
difficult to estimate using the simple models presented in this
work. The merger rate may be dependent on mass, and accurate mass
functions are needed to understand the growth of the red sequence due
to red mergers \citep[see][]{bu07}.

\subsection{Other Influences}

Although a combination of aging, quenching, and red mergers provides a
good fit to observed evolution, there may be other possible
explanation as well. One concern is the unknown dust content of
red-sequence galaxies, especially at $z\sim2.3$. The well-defined
shape of the $z\sim2.3$ red-sequence makes large dust contents
implausible. However, as these galaxies recently stopped forming
stars, they may still be in the process of losing their dust. The
red-sequence galaxies have a median best-fit $A_V$ of 0.8 mag, but the
constraints are poor with typical $1\sigma$ errors of 0.5
mag. Furthermore, as explained in \S~\ref{sec:pop_rs}, the degeneracy
with metallicity may introduce an additional systematic
error. Nonetheless, an $A_V$ of 0.8 mag would lower the $z\sim2.3$
red-sequence color in Figure~\ref{fig:dub_t} by 0.16 mag \citep[for
a][reddening law]{ca00}. The resulting evolution requires a later
formation redshift, and this results in a slightly bluer color at low
redshift. However, this effect is very small, and thus dust cannot be
primarily responsible for the slow observed color evolution.

New starbursts in red-sequence galaxies, for example triggered by
mergers, may reduce the color evolution. \cite{bi07} suggest that
quenched galaxies may undergo a second starburst due to gas
accretion. This starburst moves the galaxies back to the blue
cloud. Once the star formation is quenched for the second time, the
galaxy moves again to the red sequence. Compared to just passive
evolution, the galaxy will be bluer due to younger ages of the newly
formed stars.  In this context it is interesting to note that
\cite{la07} find that episodic star forming models provide the best
explanation of the evolution of the blue sequence.  Also, low-level
residual star formation -- or ``frosting'' of younger stars to an
older ``base'' population -- may depress the color evolution and could
explain the apparently too blue colors of the low-redshift
red-sequence galaxies \cite[see][]{tr00}. However, only $\sim$15\% of
the low-redshift field ellipticals show evidence for fairly recent
star formation \citep[e.g.,][]{yi05}, and thus this effect can not
fully explain the slow color evolution.

An evolving IMF or metallicity may also alter the rest-frame $U-B$
evolution of the red sequence. For example, in case metallicity is
lower for galaxies that form or quench at later times, $\Delta (U-B)$
will be slower. Furthermore, several authors have suggested that the
IMF of massive elliptical galaxies may be top-heavy or ``bottom
light'' \citep[see][and references therein]{vd07}. Such IMFs lead to
slower color evolution than Salpeter-like IMFs \citep{ti80}, although
the effect is relatively small \citep{vd07}.

Finally, we note that the used stellar population models
\citep{bc03,ma05} might be incomplete, and the evolution in the models
may be too strong.

\section{SUMMARY}
 
Our recent discovery of galaxies with quiescent stellar populations
beyond $z=2$ suggests that a red sequence is already in place at these
redshifts. We examined this suggestion using our NIR spectroscopic
survey of massive galaxies at $2\lesssim z\lesssim 3$. The combination
of spectroscopic redshifts and detailed continuum shapes as provided
by the NIR spectra, allows the first accurate rest-frame color and
stellar mass determinations for a massive galaxy sample beyond $z=2$.

The distribution of galaxies in the rest-frame $U-B$ color versus mass
diagram demonstrates the existence of a red sequence at $z\sim2.3$,
with a significance of $>3\sigma$. The red sequence hosts $\sim60$\%
of the stellar mass at the high mass end ($>10^{11} M_{\odot}$) at
$z\sim2.3$. We study the stellar populations of the red-sequence
galaxies using emission line diagnostics, and stellar population
modeling. The stellar continua, as provided by the NIR spectra, of
nearly all red-sequence galaxies are best-fit by specific SFRs less
than 0.05 Gyr$^{-1}$. Furthermore, in contrast to the blue galaxies,
they have no detected rest-frame optical emission lines (e.g.,
H$\alpha$), or the line emission is dominated by AGN activity. Thus,
both independent diagnostics imply that the red sequence is dominated
by galaxies with quiescent stellar populations.

By combining rest-frame $U-B$ with $B-V$, we find that the $z\sim2.3$
red-sequence galaxies are in a post-starburst phase, with typical ages
of $\sim 0.5-1$ Gyr. This finding is supported by the strong Balmer
break in the stacked spectrum of all red-sequence galaxies. Overall,
this implies that the red sequence is primarily driven by
post-starburst galaxies at this epoch, and probably has just started
to build up at $z\sim2.3$.

We study the rest-frame $U-B$ color evolution of massive galaxies by
comparing our sample with spectroscopic galaxy samples at $z\sim0.03$
and $z\sim0.73$. Remarkably, rest-frame $U-B$ evolves slowly, by only
$\sim 0.25$ mag between $z\sim2.3$ and the present. The fraction of
massive galaxies ($>10^{11} M_{\odot}$) on the red sequence increases
by only $\sim20$\% between $z\sim2.3$ and $z\sim0.0$. Similarly, the
fraction of the total stellar mass of massive galaxies on the
red-sequence increases by only $\sim15$\%. However, the number and
mass density of the massive ($>10^{11} M_{\odot}$) red-sequence
galaxies grow by factors of $\sim8$ and $\sim6$, respectively, over
the same redshift interval.

Overall, we show that the slow color evolution of the red sequence
does not allow a straightforward explanation. Simple aging models
predict a too strong color evolution, and consequently the
red-sequence galaxies are too red at $z\sim 0$. Also, such models
cannot reproduce the strong density evolution of galaxies on the red
sequence that we measure here. Presumably, the evolution is a
combination of aging, galaxy transformations and red
mergers. Furthermore, frosting of young stars, recent starbursts,
dust, and an evolving IMF may also play a minor role in the evolution
of the red sequence, but this remains to be explored.

More accurate constraints from the color evolution require independent
measurements of the metallicity, dust content, ages, current SFRs, and
IMF of red-sequence galaxies over all epochs. As for the moment, the
evolution of the number density and mass function provide the most
powerful method to study the growth of the red sequence. The slope and
spread ($\sigma_{U-B}$) can also be used to set further constraints on
its build-up. However, due to the significant errors on the rest-frame
colors and the small size of the $z\sim2.3$ spectroscopic sample, we
use neither in this work. We do note that none of the models discussed
here violate the constraints imposed by the observed scatter.

Finally, we would like to note that our findings are supported by
recent morphological studies. Using follow up high-resolution NIC2
imaging on {\em HST} and adaptive optics imaging with NIRC2 on Keck,
\cite{vd08} find very compact sizes and extremely high inferred
stellar mass densities for the red-sequence galaxies in this paper
\citep[see also][]{tr06,tr07,to07,zi07,lo07,ci08}. This finding
implies that these galaxies do not passively evolve into the
red-sequence galaxies in the local universe. Thus, both the
morphological and color evolution require active evolution to take
place that transform the $z\sim2.3$ red-sequence galaxies to those at
the current epoch.

\acknowledgments 
We thank the members of the MUSYC collaboration for their contribution
to this work. This research was supported by grants from the
Netherlands Foundation for Research (NWO), and the Leids
Kerkhoven-Bosscha Fonds. AvdW and GDI acknowledge support from NASA
grant NAG5-7697. Support from National Science Foundation grant NSF
CAREER AST-0449678 is gratefully acknowledged.

%

\end{document}